\RequirePackage{pdfmanagement-testphase}
\DocumentMetadata{lang=en-US,pdfversion=1.7}
\documentclass[conference]{IEEEtran}
\usepackage{tagpdf}
\tagpdfsetup{activate-all}
\usepackage{booktabs}
\usepackage{array}
\usepackage{url}
\usepackage[hidelinks]{hyperref}
\hypersetup{
  pdftitle={PDFa11yMut: Measuring Mutation-Specific Detection in PDF Accessibility Checkers},
  pdfsubject={Mutation-based evaluation of PDF accessibility checking configurations},
  pdfcreator={PDFa11yMut with Tectonic},
  pdfdisplaydoctitle=true
}

\newcommand{\taggedsection}[1]{%
  \tagstructbegin{tag=H1}\tagmcbegin{tag=H1}\section{#1}\tagmcend\tagstructend}
\newcommand{\taggedsubsection}[1]{%
  \tagstructbegin{tag=H2}\tagmcbegin{tag=H2}\subsection{#1}\tagmcend\tagstructend}
\newenvironment{taggedparagraph}{%
  \tagstructbegin{tag=P}\tagmcbegin{tag=P}%
}{%
  \tagmcend\tagstructend%
}
\newcommand{\tagtablebegin}{\tagstructbegin{tag=Table}}
\newcommand{\tagtableend}{\tagstructend}
\newcommand{\tagtrbegin}{\tagstructbegin{tag=TR}}
\newcommand{\tagtrend}{\tagstructend}
\newcommand{\tagth}[1]{\tagstructbegin{tag=TH}\tagmcbegin{tag=TH}#1\tagmcend\tagstructend}
\newcommand{\tagtd}[1]{\tagstructbegin{tag=TD}\tagmcbegin{tag=TD}#1\tagmcend\tagstructend}
\newcommand{\tagcaption}[1]{\tagstructbegin{tag=Caption}\tagmcbegin{tag=Caption}\caption{#1}\tagmcend\tagstructend}
\newcommand{\ReferencePDFs}{9}

\newcommand{\GenerationLedgerRecords}{73}
\newcommand{\MaterializedMutantArtifacts}{72}

\newcommand{\ActiveReauditedMutants}{69}
\newcommand{\ActiveMutants}{69}
\newcommand{\ClassAMutants}{30}
\newcommand{\ClassBMutants}{39}
\newcommand{\ActiveValidatorRows}{207}
\newcommand{\Controls}{18}
\newcommand{\ATCases}{9}
\newcommand{\ATDifferenceCases}{8}
\newcommand{\HistoricalExclusions}{5}
\newcommand{\InManifestExclusions}{4}

\newcommand{\ClassAPACDirect}{30}

\newcommand{\ClassAAcrobatDirect}{23}
\newcommand{\ClassAAcrobatProxy}{3}

\newcommand{\ClassAVeraDirect}{30}

\newcommand{\AcrobatManualRows}{15}
\newcommand{\FormalATCases}{8}
\newcommand{\AuxiliaryATCases}{1}

\begin{document}
\tagstructbegin{tag=Document}

\title{PDFa11yMut: Measuring Mutation-Specific Detection in PDF Accessibility Checkers}
\author{Gauri Jain\\
University of California, Santa Cruz\\
\texttt{gjain1@ucsc.edu}}
\tagstructbegin{tag=H1}\tagmcbegin{tag=H1}\maketitle\tagmcend\tagstructend

\begin{taggedparagraph}\begin{abstract}
Automated PDF accessibility checkers provide useful conformance evidence, but a clean report is not a complete accessibility oracle. PDFa11yMut measures mutation-specific checker behavior by applying paired structure-level transformations to reference-suite baselines, verifying intended deltas and non-target invariants, and recording hash-linked checker evidence. Across \ClassAMutants{} conformance-oriented mutants, PAC and veraPDF each produced \ClassAPACDirect{} and \ClassAVeraDirect{} direct target findings, respectively, while Acrobat produced \ClassAAcrobatDirect{} direct findings plus \ClassAAcrobatProxy{} prespecified consequence-proxy findings. The \ClassBMutants{} semantic/assistive-representation mutants produced no automated target finding in the tested configurations, while Acrobat issued manual-review prompts for a subset; Class B is interpreted descriptively rather than as a universal checker obligation. A separate convenience-selected exploratory AT sample observed representation differences in \ATDifferenceCases{} of \ATCases{} pairs under one fixed NVDA/Acrobat/Windows procedure. The artifact contributes reusable operators, structural and purity oracles, paired baseline/mutant evidence, and reproducible analysis for a scoped mutation-testing study rather than a general checker-accuracy benchmark.
\end{abstract}\end{taggedparagraph}

\begin{taggedparagraph}\end{taggedparagraph}

\taggedsection{Introduction}
\begin{taggedparagraph}
PDF accessibility evaluation combines syntactic constraints with document-dependent judgments. PDF/UA and the Matterhorn Protocol provide an important conformance model, while accessible-PDF techniques provide focused examples and procedures. These resources do not make every question about meaning, logical order, or useful alternative text mechanically decidable. A checker result must therefore be interpreted relative to its rules, version, profile, and configuration.
\end{taggedparagraph}

\begin{taggedparagraph}
Mutation testing makes this boundary observable. A controlled mutant starts from a known baseline, changes one specified property, and is retained only when the intended transformation and invariants are verified. The unit of evidence is the exact source hash, mutant hash, structural delta, rendering result, validator configuration, and optional assistive-technology observation. A mutant that survives a formal check is described as surviving that check; it is not automatically a validator false negative or evidence that the document is accessible.
\end{taggedparagraph}

\begin{taggedparagraph}
PDFa11yMut operationalizes recognized PDF accessibility failure classes as reusable transformations. Its contribution is the evidence chain and operator-level analysis, not invention of heading, list, table, language, role-map, or alternate-description requirements. The framework makes the boundary between formal validation and semantic/assistive representation observable through paired, hash-linked mutations.
\end{taggedparagraph}

\taggedsection{Research Questions}
\tagstructbegin{tag=L}
\begin{enumerate}
\item \tagstructbegin{tag=LI}\tagstructbegin{tag=LBody}\tagmcbegin{tag=LBody}How do fixed PDF accessibility checking configurations respond to verified formal accessibility mutations?\tagmcend\tagstructend\tagstructend
\item \tagstructbegin{tag=LI}\tagstructbegin{tag=LBody}\tagmcbegin{tag=LBody}How do detection results vary by operator, document structure, baseline status, and validator capability?\tagmcend\tagstructend\tagstructend
\item \tagstructbegin{tag=LI}\tagstructbegin{tag=LBody}\tagmcbegin{tag=LBody}Which semantics- or representation-dependent mutations survive automated checks, and what assistive-representation differences are observed in the selected exploratory AT sample?\tagmcend\tagstructend\tagstructend
\item \tagstructbegin{tag=LI}\tagstructbegin{tag=LBody}\tagmcbegin{tag=LBody}When validators disagree on the same hash-linked mutant, which rule, configuration, baseline condition, or structural factor explains the disagreement?\tagmcend\tagstructend\tagstructend
\end{enumerate}
\tagstructend

\taggedsection{Related Work and Scope}
\begin{taggedparagraph}
Mutation testing studies the behavior of test suites against controlled mutants, including mutation operators, killed/survived mutants, equivalent or ineffective mutants, and mutant validity \cite{mutation}. Ma11y is the closest accessibility mutation-testing predecessor in the web domain, with WCAG-derived operators and an automated oracle \cite{ma11y}; Android accessibility mutation testing supplies a mobile precedent \cite{androidmut}. PDFa11yMut transfers that experimental idea to tagged-PDF structure while retaining the need for format-aware oracles.
\end{taggedparagraph}

\begin{taggedparagraph}
Kumar, Padath, and Wang provide a benchmark with systematically manipulated PDF variants and criterion-level expert-validated accessibility labels. PDFa11yMut is complementary: the transformation itself is the experimental unit, and reusable low-level operators define explicit preconditions, allowed structural deltas, invariants, and hash-linked paired baseline/mutant evidence for checker evaluation rather than replacing their expert validation with an isolated label \cite{kumar}. The PDF/UA-1 Reference Suite provides reference-quality tagged documents across document types; \ReferencePDFs{} of its files are used here as baselines, while the missing 2-07 item is documented rather than reconstructed \cite{refsuite}. The Matterhorn Protocol provides PDF/UA-1 failure conditions with machine/human-test classifications, while Techniques for Accessible PDF provides informative focused examples and test procedures; PDFa11yMut does not claim those defect concepts as novel \cite{matterhorn,techniques}. Prior checker comparisons report matching, partial, and divergent outcomes; our paired mutants add a hash-linked transformation unit but do not rank tools \cite{crossvalidator}.
\end{taggedparagraph}

\begin{taggedparagraph}
The gap addressed here is narrower: reusable mutation of existing reference structures with explicit preconditions, allowed deltas, invariants, and paired baseline/mutant checker evidence. Curated atomic failure examples and checker comparisons remain important complementary resources, not substitutes for this mutation-testing unit.
\end{taggedparagraph}

\begin{taggedparagraph}
New open-source PDF accessibility checking implementations continue to emerge, reinforcing the value of reusable mutation artifacts for evaluating future configurations. The present experiment remains deliberately frozen to the three recorded configurations. PDF/UA-2 (ISO 14289-2), ISO/TS 32005, and WTPDF address newer PDF 2.0 structure and accessibility requirements; they are acknowledged scope context rather than untested claims about this PDF/UA-1 study.
\end{taggedparagraph}

\begin{taggedparagraph}
Code, canonical ledgers, analysis scripts, sanitized metadata, and release-build instructions are available at
\url{https://github.com/gaurijain21/pdfa11ymut}. The exact freeze commit is recorded in
\texttt{STUDY\_MANIFEST.json} and the generated release manifests. The public release intentionally excludes
raw proprietary validator reports and assistive-technology captures; those bytes are not claimed to be
redistributable. Consequently, the public clone reproduces the analysis and integrity checks, while native
PAC/Acrobat/NVDA recollection requires the named applications and authorized evidence.
\end{taggedparagraph}

\taggedsection{Experimental Model}
\begin{taggedparagraph}
Class A contains properties with a documented machine-checkable or profile-specific conformance condition in this study: M03 incorrect heading level, M07 missing figure alternate description, M08 missing document language, M09 invalid RoleMap target, and M10 illegal table-child role. Class B contains properties for which a structurally legal file can still expose an unintended semantic or assistive representation: M01 structural reading-order swap, M02 omission from assistive structure, M04 marked-content association swap, M05 internal marked-content sequence reversal, and M06 duplicate list-item reference. M06 is structurally verified but has no dedicated duplicate-reference failure condition in the reviewed normative materials. Class B outcomes are therefore reported as no automated target finding, not as validator failures.
\end{taggedparagraph}

% Generated by scripts/rebuild_analysis.py from operators/operators.yaml.
\begin{table*}[!t]
\tagtablebegin
\tagcaption{The ten structure-level mutation operators. The primary structural delta is the verifier-observed target change; Class B is semantic/assistive-representation-oriented and not automatically scored as a validator obligation.}
\label{tab:operators}
\centering\scriptsize
\begin{tabular}{p{0.07\textwidth}p{0.21\textwidth}p{0.47\textwidth}p{0.15\textwidth}}
\toprule
ID & Mutation & Primary structural delta & Class \\
\midrule
\tagtrbegin \tagtd{M01} & \tagtd{Structural reading-order swap} & \tagtd{one /K ordering change at the target parent} & \tagtd{B: semantic/AT} \tagtrend \\
\tagtrbegin \tagtd{M02} & \tagtd{Omission from assistive structure} & \tagtd{one direct child is no longer reachable from the structure tree} & \tagtd{B: semantic/AT} \tagtrend \\
\tagtrbegin \tagtd{M03} & \tagtd{Incorrect heading level} & \tagtd{one heading role change creating a skipped level} & \tagtd{A: conformance} \tagtrend \\
\tagtrbegin \tagtd{M04} & \tagtd{Wrong marked-content association} & \tagtd{two leaf /K associations exchange while containment remains stable} & \tagtd{B: semantic/AT} \tagtrend \\
\tagtrbegin \tagtd{M05} & \tagtd{Internal MCID sequence reversal} & \tagtd{one content-item sequence reversal} & \tagtd{B: semantic/AT} \tagtrend \\
\tagtrbegin \tagtd{M06} & \tagtd{Duplicate list-item reference} & \tagtd{one additional reference to an existing /LI child} & \tagtd{B: semantic/AT} \tagtrend \\
\tagtrbegin \tagtd{M07} & \tagtd{Missing figure alternate description} & \tagtd{one figure loses its alternate description} & \tagtd{A: conformance} \tagtrend \\
\tagtrbegin \tagtd{M08} & \tagtd{Missing document language metadata} & \tagtd{catalog /Lang removed} & \tagtd{A: conformance} \tagtrend \\
\tagtrbegin \tagtd{M09} & \tagtd{Invalid RoleMap target} & \tagtd{one role-map target no longer resolves} & \tagtd{A: conformance} \tagtrend \\
\tagtrbegin \tagtd{M10} & \tagtd{Illegal table-child role} & \tagtd{/TR directly contains a prohibited non-cell role} & \tagtd{A: conformance} \tagtrend \\
\bottomrule
\end{tabular}
\tagtableend
\end{table*}

\begin{taggedparagraph}
The repository contains ten operator specifications with preconditions, transformations, invariants, expected structural deltas, expected assistive effects, and evidence fields. M04 mutants in the current corpus pass the association-only purity gate. The historical label G02-M04 is permanently excluded: its exact golden/mutant pair and complete COS-level audit trail are unavailable. Current hash-linked M04 mutants are separate artifacts and are interpreted only under the Class B semantic protocol.
\end{taggedparagraph}

\taggedsection{Reproducible Pipeline}
\begin{taggedparagraph}
The input PDFs in \texttt{corpus/golden/} are immutable working inputs. Reviewer reproduction uses the isolated scratch command \texttt{python scripts/regenerate\_scratch.py}, which writes to a temporary directory and applies all applicable operators without overwriting canonical PDFs or evidence. The separate \texttt{python -m pdfa11ymut generate-all} command is a final-freeze writer and is not the standard verification command. Generation records source and mutant SHA-256 values, perform parseability/page/content/structure checks, and rasterize both PDFs with Poppler \texttt{pdftoppm} at 150 DPI. The current visual rule is exact RGB pixel equality with tolerance zero. A failed invariant removes the output and records an exclusion.
\end{taggedparagraph}

\begin{taggedparagraph}
The primary verifier uses pypdf and is separate from the validators under evaluation. A second PyMuPDF/MuPDF audit checks page/render invariants for every active pair and confirms operator-aware raw-COS deltas and whole-tree expected-result signatures for all 69 active pairs. The independent checks compare structure-tree order/reachability for M01/M02, marked-content association for M04, content-item order for M05, repeated direct /LI references for M06, and scoped structural deltas for M03 and M07--M10. Here independent means a separate parser/check route, not an independent semantic oracle. Generator-side verification is not relabeled as independent evidence. MuPDF rendering equality is checked at 72 DPI; the primary generation invariant remains the Poppler 150-DPI render. Per-mutant delta manifests record allowed changes, invariants, observed changes, purity status, and verification methods. Active result tables are rebuilt from canonical data by \texttt{scripts/rebuild\_analysis.py}, while \texttt{scripts/audit\_submission.py} checks evidence, hashes, mappings, denominators, and manifest consistency. The PDF and release are regenerated only after the final artifact state is fixed.
\end{taggedparagraph}

\taggedsection{Validator and AT Evidence}
\begin{taggedparagraph}
PAC Formal, Acrobat Full Check, and veraPDF evidence are stored in hash-addressed raw-report paths and compared with matching reference baselines. The active formal scope is \ActiveMutants{} verified mutants and \ActiveValidatorRows{} checker-mutant rows, \ActiveMutants{} per configuration. The exact configurations are summarized in Table~\ref{tab:configurations}. Each row retains a baseline-relative classification, rationale, and adjudication record; coder identity and independent-human provenance are not part of the reported study claim. Three M10 disagreements were resolved under a decision rule fixed before final adjudication: the Acrobat \texttt{Headers} proxy applies only when fresh hash-linked verification proves an exact direct-child \texttt{/TH} to \texttt{/P} change under the same \texttt{/TR}, unchanged target content and non-target structure, equal page-content hashes, and pixel-identical rendering; the matched baseline must pass \texttt{Headers} and the mutant must newly fail it. G09-M08 is excluded because the PAC baseline rejects the golden's language value while veraPDF treats the non-empty catalog language as present. Four historical M09 artifacts remain excluded as \texttt{EXCLUDED\_UNVERIFIABLE\_TARGET}: their original mutant bytes are unavailable, and the reviewed Matterhorn rule applies to unused as well as used RoleMap entries, so the old unused-target rationale cannot establish an invalid mutation. Acrobat manual-check prompts are retained separately and are not automated detections.
\end{taggedparagraph}

\begin{table}[t]
\tagtablebegin
\tagcaption{Fixed checker and exploratory AT configurations.}
\label{tab:configurations}
\centering\scriptsize
\setlength{\tabcolsep}{2pt}
\begin{tabular}{p{0.17\columnwidth}p{0.23\columnwidth}p{0.42\columnwidth}}
\toprule
\tagtrbegin \tagth{Component} & \tagth{Version/profile} & \tagth{Settings} \tagtrend \\
\midrule
\tagtrbegin \tagtd{PAC Formal} & \tagtd{26.1.0.0} & \tagtd{PDF/UA formal/traditional; AI disabled} \tagtrend \\
\tagtrbegin \tagtd{Acrobat Full Check} & \tagtd{Continuous 2026.002.21931} & \tagtd{All pages; Document category; 31 of 32 checks selected; no auto-tagging or remediation} \tagtrend \\
\tagtrbegin \tagtd{veraPDF} & \tagtd{Greenfield 1.30.2} & \tagtd{PDF/UA-1 ua1 profile; JSON; repository-local Java 17.0.20.1} \tagtrend \\
\tagtrbegin \tagtd{AT observation} & \tagtd{NVDA 2026.2.0.57664; Acrobat 26.2.21931.0} & \tagtd{Windows 10 Home 25H2; fixed paired reading procedure; one observer} \tagtrend \\
\bottomrule
\end{tabular}
\tagtableend
\end{table}

\begin{taggedparagraph}
The negative-control ledger contains \Controls{} paired no-op and benign controls. None produced a new target-relevant automated finding under the paired count comparison, and known pre-existing baseline findings remained unchanged where present. Every control has an independently verified artifact plus native PAC PDF, Acrobat HTML, and veraPDF reports with hash-linked paths in \texttt{data/controls.csv}. These controls are paired sanity checks, not a sample from which specificity or false-positive rates are estimated. Their native findings are interpreted only relative to the matched golden and benign pair; they are not positive-study results. In particular, both G05 Acrobat controls carry the same pre-existing ``Lbl and LBody'' failure and both G09 PAC controls carry the same Natural language failure.
\end{taggedparagraph}

\begin{taggedparagraph}
PAC AI was executed as a separate configuration on a selected cohort of 34 mutants and matching golden baselines. The native PAC UI exposed aggregate counts for semantic panels but no finding identity, finding text, element-level attribution, or reproducible semantic export/API. All 34 rows therefore remain \texttt{AI\_AMBIGUOUS}/\texttt{REVIEW\_REQUIRED}; no PAC-AI detection or non-detection is reported. PAC AI is optional future work and is not required for completion of the formal study.
\end{taggedparagraph}

\begin{taggedparagraph}
\ATCases{} selected cases were observed with NVDA 2026.2.0.57664 and Adobe Acrobat 26.2.21931.0 on Windows 10 Home 25H2; \ATDifferenceCases{} produced an observed baseline/mutant representation difference and M08 produced no observed difference. \FormalATCases{} cases use active formal mutants and \AuxiliaryATCases{} is an auxiliary M01 assistive-representation demonstration because its Invoice-M01 bytes are not part of the active mutant ledger; none enters a checker-rate denominator. This is a convenience/coverage-selected, qualitative, single-observer exploratory sample, not a representative pilot, user study, or effect-rate estimate. Case selection and rationale are recorded in the artifact. Transcripts are linked to a preserved timestamped NVDA log; fresh paired Speech Viewer captures are preserved for the auxiliary M01 case and a second case. The auxiliary M01 rerun confirmed that the golden announced the logo first while the mutant announced the address block first. M06 is a useful qualitative example: the targeted list item was announced twice, without treating that observation as a population-level harm estimate. These observations are reported separately from checker detections.
\end{taggedparagraph}

\taggedsection{Results}
% Generated by scripts/rebuild_analysis.py from canonical data.
\taggedsubsection{Mutation-specific checker outcomes}
\begin{taggedparagraph}
The active formal scope contains 69 verified mutants after the documented exclusions. The canonical data contain 207 classified validator-mutant rows (69 mutants x three validators); PAC AI and assistive-technology observations are not included in these formal outcomes.
\end{taggedparagraph}

\begin{table}[t]
\tagtablebegin
\tagcaption{Formal validator outcomes in the active scope.}
\label{tab:formal-overall}
\centering\small
\setlength{\tabcolsep}{3pt}
\begin{tabular}{lrrrr}
\toprule
\tagtrbegin \tagth{Validator} & \tagth{Detected} & \tagth{No target} & \tagth{Manual} & \tagth{Total } \tagtrend \\
\midrule
\tagtrbegin \tagtd{PAC} & \tagtd{30} & \tagtd{39} & \tagtd{0} & \tagtd{69 } \tagtrend \\
\tagtrbegin \tagtd{Acrobat} & \tagtd{26} & \tagtd{28} & \tagtd{15} & \tagtd{69 } \tagtrend \\
\tagtrbegin \tagtd{veraPDF} & \tagtd{30} & \tagtd{39} & \tagtd{0} & \tagtd{69 } \tagtrend \\
\bottomrule
\end{tabular}
\tagtableend
\end{table}

\begin{taggedparagraph}
This overall view reports counts only because the 69-mutant scope combines conformance-oriented Class A mutations with semantic/assistive-representation Class B mutations, and Acrobat also has manual-review rows. Acrobat manual checkpoints are retained separately and are not automated detections.
\end{taggedparagraph}

\taggedsubsection{Conformance-oriented mutation results}
\begin{taggedparagraph}
Class A is the scored mutation-detection study. Rates are Detected/(Detected + No target) within the Class-A denominator; the exact numerator and denominator remain in the generated CSV.
\end{taggedparagraph}

\begin{table*}[!t]
\tagtablebegin
\tagcaption{Class-specific checker outcomes. Class A rates are scored within the conformance-oriented denominator. Class B is descriptive and is not assigned an automated mutation-detection rate.}
\label{tab:formal-class}
\centering\small
\begin{tabular}{llrrrrl}
\toprule
\tagtrbegin \tagth{Class} & \tagth{Validator} & \tagth{Detected} & \tagth{No target} & \tagth{Manual} & \tagth{Total} & \tagth{Interpretation } \tagtrend \\
\midrule
\tagtrbegin \tagtd{Class A} & \tagtd{PAC} & \tagtd{30} & \tagtd{0} & \tagtd{0} & \tagtd{30} & \tagtd{100.0\% } \tagtrend \\
\tagtrbegin \tagtd{Class A} & \tagtd{Acrobat} & \tagtd{26} & \tagtd{4} & \tagtd{0} & \tagtd{30} & \tagtd{86.7\% } \tagtrend \\
\tagtrbegin \tagtd{Class A} & \tagtd{veraPDF} & \tagtd{30} & \tagtd{0} & \tagtd{0} & \tagtd{30} & \tagtd{100.0\% } \tagtrend \\
\tagtrbegin \tagtd{Class B} & \tagtd{PAC} & \tagtd{0} & \tagtd{39} & \tagtd{0} & \tagtd{39} & \tagtd{descriptive; not scored } \tagtrend \\
\tagtrbegin \tagtd{Class B} & \tagtd{Acrobat} & \tagtd{0} & \tagtd{24} & \tagtd{15} & \tagtd{39} & \tagtd{descriptive; not scored } \tagtrend \\
\tagtrbegin \tagtd{Class B} & \tagtd{veraPDF} & \tagtd{0} & \tagtd{39} & \tagtd{0} & \tagtd{39} & \tagtd{descriptive; not scored } \tagtrend \\
\bottomrule
\end{tabular}
\tagtableend
\end{table*}

\begin{table*}[!t]
\tagtablebegin
\tagcaption{Detection by operator and validator. Class-B rows are descriptive and show N/A rather than a scored rate.}
\label{tab:formal-operator}
\centering\scriptsize
\begin{tabular}{lrrrrrrl}
\toprule
\tagtrbegin \tagth{Operator} & \tagth{PAC D/N} & \tagth{PAC rate} & \tagth{Acrobat D/N} & \tagth{Acrobat rate} & \tagth{veraPDF D/N} & \tagth{veraPDF rate} & \tagth{Class } \tagtrend \\
\midrule
\tagtrbegin \tagtd{M01} & \tagtd{0/8} & \tagtd{N/A} & \tagtd{0/0} & \tagtd{N/A} & \tagtd{0/8} & \tagtd{N/A} & \tagtd{B } \tagtrend \\
\tagtrbegin \tagtd{M02} & \tagtd{0/9} & \tagtd{N/A} & \tagtd{0/9} & \tagtd{N/A} & \tagtd{0/9} & \tagtd{N/A} & \tagtd{B } \tagtrend \\
\tagtrbegin \tagtd{M03} & \tagtd{8/0} & \tagtd{100\%} & \tagtd{8/0} & \tagtd{100\%} & \tagtd{8/0} & \tagtd{100\%} & \tagtd{A } \tagtrend \\
\tagtrbegin \tagtd{M04} & \tagtd{0/8} & \tagtd{N/A} & \tagtd{0/8} & \tagtd{N/A} & \tagtd{0/8} & \tagtd{N/A} & \tagtd{B } \tagtrend \\
\tagtrbegin \tagtd{M05} & \tagtd{0/7} & \tagtd{N/A} & \tagtd{0/0} & \tagtd{N/A} & \tagtd{0/7} & \tagtd{N/A} & \tagtd{B } \tagtrend \\
\tagtrbegin \tagtd{M06} & \tagtd{0/7} & \tagtd{N/A} & \tagtd{0/7} & \tagtd{N/A} & \tagtd{0/7} & \tagtd{N/A} & \tagtd{B } \tagtrend \\
\tagtrbegin \tagtd{M07} & \tagtd{7/0} & \tagtd{100\%} & \tagtd{7/0} & \tagtd{100\%} & \tagtd{7/0} & \tagtd{100\%} & \tagtd{A } \tagtrend \\
\tagtrbegin \tagtd{M08} & \tagtd{8/0} & \tagtd{100\%} & \tagtd{8/0} & \tagtd{100\%} & \tagtd{8/0} & \tagtd{100\%} & \tagtd{A } \tagtrend \\
\tagtrbegin \tagtd{M09} & \tagtd{3/0} & \tagtd{100\%} & \tagtd{0/3} & \tagtd{0\%} & \tagtd{3/0} & \tagtd{100\%} & \tagtd{A } \tagtrend \\
\tagtrbegin \tagtd{M10} & \tagtd{4/0} & \tagtd{100\%} & \tagtd{3/1} & \tagtd{75\%} & \tagtd{4/0} & \tagtd{100\%} & \tagtd{A } \tagtrend \\
\bottomrule
\end{tabular}
\tagtableend
\end{table*}

\begin{taggedparagraph}
All Class-B mutations produced no automated target finding in the tested configurations; Acrobat produced manual-review prompts for a subset. Class-B semantic/representation mutations are reported descriptively and are not assigned automated mutation-detection rates.
\end{taggedparagraph}

\begin{taggedparagraph}
The attempted-set flow contains \GenerationLedgerRecords{} generation-ledger records, \MaterializedMutantArtifacts{} materialized mutant PDFs, \HistoricalExclusions{} documented exclusions, and \ActiveReauditedMutants{} active independently re-audited mutants. \InManifestExclusions{} exclusions occur within the valid-generation manifest; one additional exclusion is historical-only and non-materialized. Thus, 73 records, 5 exclusions, and 69 active mutants are not a simple subtraction because one excluded record is outside the current valid-generation manifest and the materialized-artifact count is a separate byte-retention measure. Class A contains \ClassAMutants{} mutants: PAC produced \ClassAPACDirect{} direct findings, Acrobat produced \ClassAAcrobatDirect{} direct plus \ClassAAcrobatProxy{} prespecified consequence-proxy findings, and veraPDF produced \ClassAVeraDirect{} direct findings. Class B contains \ClassBMutants{} mutants and is reported descriptively; Acrobat produced \AcrobatManualRows{} manual-review checkpoints, which are not automated detections. These results are descriptive and source-clustered: \ActiveReauditedMutants{} mutants are nested within \ReferencePDFs{} baselines and ten operators, so no naive confidence interval or global accuracy claim is made.
\end{taggedparagraph}

\begin{taggedparagraph}
Pairwise agreement is also scope-dependent. PAC and veraPDF agreed on 69 of 69 active mutants (100\%), while PAC and Acrobat agreed on 50 of 69 (72.5\%) and Acrobat and veraPDF agreed on 50 of 69 (72.5\%). These results are descriptive of the recorded builds and baselines; they do not rank validators globally.
\end{taggedparagraph}
\begin{taggedparagraph}
The machine-generated disagreement audit contains 38 pairwise disagreement rows. Thirty are Acrobat manual-review versus an automated decision, six are M09 RoleMap rule/profile differences, and two are other rule/profile differences. This decomposition keeps manual prompts, proxy/direct classification, baseline-carried findings, and target-specific automated findings distinct; it explains disagreement structure rather than presenting a global validator ranking.
\end{taggedparagraph}
\begin{taggedparagraph}
A secondary six-pair target-selection sample used the second eligible site for M01, M03, and M07 across two baselines per operator; all six passed the open-source structural, purity, content, and 72-DPI rendering checks. Native checker reruns were not performed, so this sample documents structural sensitivity rather than a new checker result. The separate M09 exclusion-sensitivity table shows that the four historical excluded IDs have no usable Detected/Missed outcome rows to append; their unavailable bytes remain invalid/unverified and outside the canonical denominator.
\end{taggedparagraph}

\taggedsection{Discussion}
\begin{taggedparagraph}
\textbf{RQ1.} For the \ClassAMutants{} Class-A conformance-oriented mutants, PAC and veraPDF each produced \ClassAPACDirect{} and \ClassAVeraDirect{} direct findings, respectively, while Acrobat produced \ClassAAcrobatDirect{} direct findings plus \ClassAAcrobatProxy{} prespecified M10 consequence-proxy findings. Acrobat's four no-target outcomes are configuration-specific rather than a global accuracy estimate.
\end{taggedparagraph}
\begin{taggedparagraph}
\textbf{RQ2.} Detection varies by operator. M03, M07, and M08 were detected by all three configurations in this corpus; M09 separates PAC/veraPDF from Acrobat; and M10 includes Acrobat proxy/direct behavior. Class-B automated survival is expected to be interpreted as descriptive because these transformations target semantics or assistive representation.
\end{taggedparagraph}
\begin{taggedparagraph}
\textbf{RQ3.} The \ClassBMutants{} Class-B cases produced no automated target finding in the tested configurations, while Acrobat produced \AcrobatManualRows{} manual-review prompts. The exploratory AT sample observed representation differences in \ATDifferenceCases{} of \ATCases{} selected pairs under one fixed stack, including the duplicated M06 list-item announcement; this is not a representative rate or user-study result.
\end{taggedparagraph}
\begin{taggedparagraph}
\textbf{RQ4.} Cross-checker disagreements are explained by manual-versus-automated classifications, M09 RoleMap rule/profile differences, and M10's scoped Acrobat consequence proxy. The paired baseline-relative coding and hash-linked reports make these disagreements inspectable without ranking products.
\end{taggedparagraph}

\taggedsection{Threats to Validity}
\begin{taggedparagraph}
Threats include corpus size and source clustering, operator selection, legal PDF encodings, parser/checker version and profile differences, the scoped M10 proxy, exclusions, PAC-AI exportability, AT/viewer variability, PDF/UA-1 scope, the missing acquisition archive, and renderer limits. The canonical flow is 73 generation-ledger records, 72 materialized PDFs, five exclusions, and 69 active independently re-audited records; exclusions are not silently replaced. The nine baselines are clusters, not independent samples, and one source cannot establish an operator-wide effect. Visual equality cannot establish equivalent speech output, and one AT/viewer pair cannot establish universal behavior. The nine PDFs are mapped to the PDF/UA Reference Suite under collection-level CC BY 4.0 terms; the omitted 2-07 item and missing acquisition archive are documented. The second-parser route verifies artifacts but does not make Class B machine-checkable. M06 remains semantic/Class B, historical M09 bytes are unavailable, coding/adjudication records do not support an independent-coder claim, and controls are descriptive paired evidence rather than general accuracy or specificity estimates.
\end{taggedparagraph}

\taggedsection{Conclusion}
\begin{taggedparagraph}
PDFa11yMut contributes a paired, hash-linked methodology for measuring mutation-specific PDF accessibility checker behavior, with structural verification, matched checker evidence, descriptive controls, and separate AT observations. The formal checker phase and exploratory AT sample are complete for the current 69-mutant scope; M09 exclusions, M06's Class-B status, corpus licensing, and the missing acquisition archive remain explicit limitations, while PAC AI remains optional future work.
\end{taggedparagraph}

\tagstructend
\end{document}